\documentclass[10pt,conference]{IEEEtran}
\IEEEoverridecommandlockouts
\usepackage{cite}
\usepackage{amsmath,amssymb,amsfonts}
\usepackage{algorithmic}
\usepackage{graphicx}
\usepackage{textcomp}
\usepackage{xcolor}
\def\BibTeX{{\rm B\kern-.05em{\sc i\kern-.025em b}\kern-.08em
    T\kern-.1667em\lower.7ex\hbox{E}\kern-.125emX}}

\usepackage{xurl}
\usepackage{booktabs}
\usepackage{arydshln}
\usepackage{balance}
\usepackage{pdfpages}

\begin{document}

\title{Distributed-Ledger-based Authentication with\\Decentralized Identifiers and Verifiable Credentials}

\author{
\IEEEauthorblockN{Zolt\'{a}n Andr\'{a}s Lux}
\IEEEauthorblockA{\textit{Service-centric Networking} \\
\textit{Technische Universit{\"a}t Berlin}\\
Berlin, Germany \\
zoltan.a.lux@tu-berlin.de}
\and
\IEEEauthorblockN{Dirk Thatmann}
\IEEEauthorblockA{\textit{Telekom Innovation Laboratories} \\
\textit{Deutsche Telekom}\\
Berlin, Germany \\
dirk.thatmann@telekom.de}
\and
\IEEEauthorblockN{Sebastian Zickau}
\IEEEauthorblockA{\textit{Service-centric Networking} \\
\textit{Technische Universit{\"a}t Berlin}\\
Berlin, Germany \\
sebastian.zickau@tu-berlin.de}
\and
\IEEEauthorblockN{Felix Beierle}
\IEEEauthorblockA{\textit{Service-centric Networking} \\
\textit{Technische Universit{\"a}t Berlin}\\
Berlin, Germany \\
beierle@tu-berlin.de}
}

\IEEEoverridecommandlockouts
\IEEEpubid{\makebox[\columnwidth]{   \hfill} \hspace{\columnsep}\makebox[\columnwidth]{ }}

\maketitle

\IEEEpubidadjcol

\begin{abstract}
Authentication with username and password is becoming an inconvenient process for the user.
End users typically have little control over their personal privacy,
and data breaches effecting millions of users have already happened several times.
We have implemented a proof of concept decentralized OpenID Connect Provider by marrying it with Self-Sovereign Identity, which gives users the freedom to choose from a very large pool of identity providers instead of just a select few corporations, thus enabling the democratization of the highly centralized digital identity landscape. Furthermore, we propose a verifiable credential powered decentralized Public Key Infrastructure using distributed ledger technologies, which creates a straightforward and verifiable way for retrieving digital certificates.
\end{abstract}

\begin{IEEEkeywords}
authentication, self-sovereign identity, verifiable credentials, distributed ledger technologies
\end{IEEEkeywords}

\section{Introduction}

Many websites support Facebook and Google single sign-on (SSO) solutions for end users.
Unfortunately, this induces a cost in terms of privacy.
Often, the business interests of such centralized service providers like Facebook and Google are not aligned with the users' interests.
This may lead to the users' data being used beyond their intention when initially sharing it \cite{FalchBusinessmodelssocial2009}.
Users could still seek the benefits of the SSO customer experience of not having to register at each web site, without sacrificing their privacy.

The idea of Self-Sovereign Identity (SSI) aims at a solution for this by decentralizing digital identities. The W3C standardization efforts on \textit{verifiable credentials} and \textit{decentralized identifiers} serve this vision of SSI. Currently, distributed ledger technology (DLT), such as Sovrin and Hyperledger Indy, is favored for implementing SSI frameworks. With these approaches, end users gain more privacy and more control over their data, yet are able to have more trust online than we have today.
NIST defines authentication as \textit{verifying the identity of a user, process, or device} and authorization as \textit{the right or a permission that is granted to a system entity to access a system resource} \cite{NistNISTSpecialPublication2012} \cite{StoufferGuideIndustrialControl}.

The industry standards behind SSO solutions are OAuth 2.0 and OpenID Connect (OIDC) \cite{sakimura2014openid}.
OIDC is an authentication layer on top of OAuth 2.0 \cite{jones2012oauth2}, an authorization framework, as defined on the official OIDC website\footnote{\url{https://openid.net/connect/}}. By default, there is an OIDC Provider such as Facebook or Google providing information about the identity holder towards an OIDC Client.
A self-sovereign version of OIDC is to ask the user for the personal information without actually having a user account at the SSI OIDC Provider.
Thus, there is no central database with personal data that the user has to rely on and that is prone to be hacked.

Additionally, there are valid concerns about the security of existing authentication systems.
Even certain two factor authentication methods are not secure enough for use cases like online banking in 2019\footnote{\url{https://en.secnews.gr/182810/phishing-taftotita-parakampsi/}}, since powerful frameworks like Muraena and NecroBrowser are able to attack time-based one-time passwords (TOTP) generated via software (e.g., Google Authenticator) or hardware (e.g., RSA SecurID).
Furthermore, the widely used SMS tokens can be intercepted with SIM swap attacks and used for replay attacks \cite{MullinerSMSBasedOneTimePasswords2013}.
Data breaches effecting millions of users already happened even at companies like Facebook and Google, and end users typically have little control over their data.

SSI can give the control back into the end user's hands and potentially help alleviate those issues of privacy concerns, security, and data breaches.
When combining SSI and OIDC for user authentication,
numerous possible approaches could be followed. 
In this paper, we perform a thorough analysis for finding a suitable method for our scenario.

SSI can facilitate user authentication and authorization in multiple scenarios.
Besides OIDC, we can also use SSI technology for integrating with, extending, and modifying existing digital identity standards like X.509 \cite{myers1999x}, and thus delivering enhanced interoperability.
There have been several initiatives aiming to establish and improve online identification processes. The Domain Name System (DNS) and Public Key Infrastructure (PKI) rely on X.509 and the centralized system of Certificate Authorities (CA) enabling websites to authenticate themselves toward people browsing the Internet. Unfortunately, the current PKI can induce high certificate costs, and CA malfunctioning, like in the case of DigiNotar, can cause tremendous problems.

The main contributions of this paper are (1) reviewing SSI authentication methods, (2) implementing and evaluating self-sovereign authentication for OIDC, and (3) analyzing the benefits of an SSI-powered PKI.

In the following,
we start by giving background information on SSI, decentralized identifiers, and verifiable contracts in Section \ref{sec:background}.
In Section \ref{sec:relatedwork}, we survey related work in the field of user authentication and PKIs utilizing distributed ledgers.
In Section \ref{sec:conceptdesign}, we present an overview about the design of our approach.
Section \ref{sec:implementation} gives an overview about the implementation
and in Section \ref{sec:evaluation}, we evaluate our approach
before concluding in Section \ref{sec:conclusion}.

\section{Background}
\label{sec:background}

Decentralizing the currently highly centralized digital identity ecosystem to deliver more
control, privacy, and security
for the end users with blockchain has been on the agenda of a select subgroup of startups, large organizations, and research institutions alike.

SSI has started back in 2016 with the blog entry of Christopher Allen\footnote{\url{http://www.lifewithalacrity.com/2016/04/the-path-to-self-soverereign-identity.html}}.
Allen proposes 10 principles guiding the design of identity systems focusing on the benefit of end users serving as the ideological foundations of SSI. The 10 principles, \textit{existence, control, access, transparency, persistence, portability, interoperability, consent, minimalization}, and \textit{protection}, formulated by Allen emphasize privacy, independence, and rights of the identity owner, also called identity holder, alongside the interoperability and the transparency of identity management systems. Furthermore, Allen is also a proponent of decentralization. Examining the current identity ecosystem, we could even deem Allen's vision of self-sovereignty utopian. There are startups, institutions, and foundations intending to decentralize digital identity, such as the European self-sovereign identity framework (eSSIF). The members of the Decentralized Identity Foundation\footnote{\url{https://identity.foundation/}} (DIF) include IBM, Microsoft, Mastercard, Sovrin, evernym, Jolocom, uPort, Auth0, the Ministry of Citizen's Services in British Columbia, and others.

\subsection{Decentralized identifiers}
A Decentralized Identifier (DID) \cite{did2019w3c} provides a verifiable and decentralized means for interacting with a DID Subject controlling the DID. A DID can be resolved to a DID Document, which can contain cryptographic material, verification methods, and service endpoints. An example DID is:

\begin{center}
\texttt{did:sov:WRfXPg8dantKVubE3HX8pw}    
\end{center}

\noindent where \texttt{did} tells us that it is a DID, \texttt{sov} is the DID Method Name for Sovrin DIDs, and \texttt{WRfXPg8dantKVubE3HX8pw}

\noindent identifies the DID subject. A DID by itself is not designed to provide trusted personal information about the DID subject on its own but to enable self-sovereignty of the holder and facilitate privacy.

For example, with Sovrin, an identity holder like Jane Doe is expected to have a different DID for each connection. As Jane has a different DID when interacting with her bank from the one she uses with an online shop, the bank and the online shop cannot use Jane's DID to correlate personal information collected about her.

\subsection{Verifiable credentials}

DIDs let users authenticate themselves online in a privacy-preserving and decentralized manner but in many use cases we need to obtain trusted information about one's identity to carry out certain transactions, such as buying alcoholic beverages online. When Jane is ordering her favorite wine in an online store, she also needs to prove that she is legally an adult. This can be achieved by using Verifiable Credentials (VC) which complement DIDs by providing means for receiving trusted identity information from end users, which is cryptographically verifiable. 

In general, VC \cite{vc2019w3c} provide us with a digital equivalent of credentials we use in our daily lives like a driver's license, a passport, or a university degree in a secure, privacy-preserving, and machine-verifiable manner.

VCs support selective disclosure, so end users can prove claims about their identity without revealing more information than they intend and need for performing a specific action. For example, when Jane orders her preferred wine in an online wine store, she only proves to the seller that she is older than 18 (assuming 18 to be the legal drinking age), which can be achieved by generating a proof about this date of birth using her passport VC. Furthermore, Jane does not even have to share the exact date, as it is sufficient for her to generate a zero-knowledge proof stating that she is older than 18.

\begin{figure*}[t!]
	\centering
    \includegraphics[width=\textwidth]{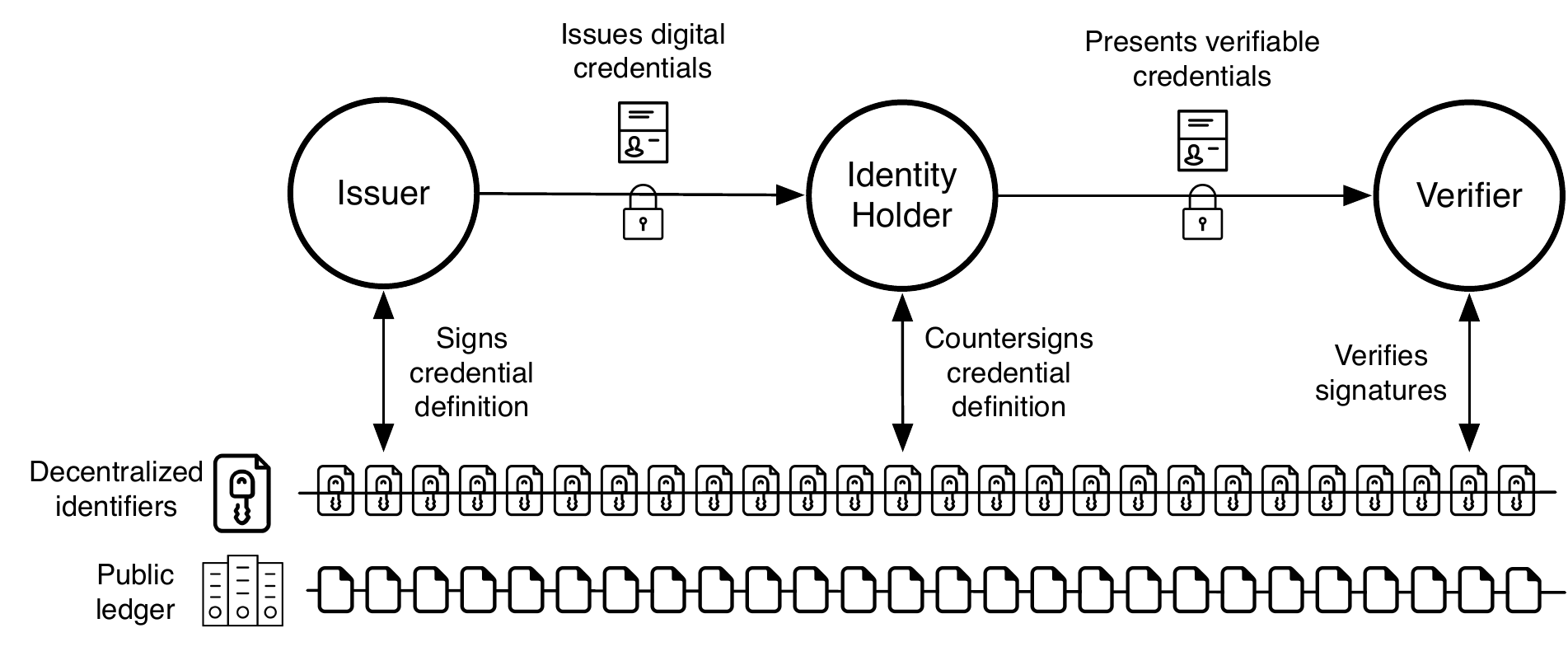}
	\caption{The Verifiable Credential (VC) role model includes an issuer, an identity holder, and a verifier. In addition, there is a publicly readable verifiable data registry, which can be a blockchain, a distributed ledger, or any secure decentralized storage. For example, a municipal office (issuer) issues a passport credential to Jane Doe (identity holder), who then presents it to her bank (verifier), when opening a new account.}
	\label{fig:ssi_role_model}
\end{figure*}

VCs can express any information that a physical credential contains, but the usage of digital signatures from both the issuer and holder make them tamper-evident and more trustworthy to the verifier. We show an illustration of the VC data model in Figure \ref{fig:ssi_role_model}.

\section{Related Work}
\label{sec:relatedwork}

In this section, we discuss online authentication of end users using OIDC and domains with the PKI.

\subsection{User authentication}

Authentication is often discussed together with authorization. OAuth 2.0 is the current industry standard authorization protocol, and also serves as the foundation of OIDC. OAuth 2.0 enables applications to gain access to already existing accounts at another HTTP web service such as Facebook. The authentication is thereby delegated to a third party that is already in contact with the user. There are 4 roles in the OAuth 2.0 model, namely (1) resource owner, (2) client, (3) resource server, and (4) authorization server.

OIDC is then placed on top of the OAuth 2.0 authorization layer in order to establish a standardized way of receiving information about user accounts from identity providers (IdPs).

VCAuthN for OIDC\footnote{\url{https://github.com/bcgov/vc-authn-oidc}} is a project from the government of British Columbia using the DIDComm protocol between the identity holder and the OIDC Provider. The OIDC Provider takes the verifier role in the SSI terminology. DIDComm is currently being incubated under the Hyperledger Aries project\footnote{https://github.com/hyperledger/aries}. The VCAuthN project integrates with both the implicit and authorization code flows. VCAuthN relies on VC presentations to extract attributes for the ID token containing user information. The VC presentations are presented either through a QR code or a deep link to the identity wallet mobile app.

However, user authentication can be approached with a focus on DIDs as well. The DID Auth document \cite{Github:DIDAuth} written by Sabadello et al. describes 10 different architectures for DID-based authentication and relations of DID Auth to other identity technologies like biometrics, OpenPGP \cite{elkins2001mime}, WebAuthn \cite{brand2019webauth}, and OIDC. The authors propose DID Auth as a possible local authentication method for an OIDC Provider and as an alternative OIDC provider discovery method through the \texttt{service endpoint} attribute of the DID Document obtained through DID resolution.

The did-auth-jose library\footnote{\url{https://github.com/decentralized-identity/did-auth-jose}} hosted in the GitHub repository of the DIF provides decentralized authentication with Javascript Object Signing and Encryption\footnote{\url{https://jose.readthedocs.io/en/latest/}} (JOSE) following the OIDC authentication flow with self-issued ID tokens. The did-auth-jose library needs access to a Universal Resolver\footnote{\url{https://uniresolver.io/}} instance for obtaining the cryptographic keys of DIDs for authentication.

The DIF has also created a specification of DID authentication for OIDC\footnote{\url{https://github.com/decentralized-identity/did-siop}}. 
In this variant, the OIDC Client does not rely on a preconfigured OIDC Provider but uses a Self-issued OIDC Provider (SIOP). The proposed SIOP DID AuthN method does not utilize the Authorization Code Flow, as integrating DID Auth with an OIDC Provider is deemed unwanted for achieving maximum decentralization and privacy. 
On the other hand, requiring OIDC Clients that implement the SIOP flow makes it more difficult to adapt SIOP DID AuthN. Nevertheless, SIOP is part of the OIDC specification.

Gr\"{u}ner, M\"{u}hle, and Meinel have also proposed a possible way for integrating OIDC with two SSI platforms \cite{gruner19ssioidc}, namely Jolocom and uPort. The architecture of their solution includes a trust engine component, which determines whether the issuer of a credential is trustworthy. This is a notable difference from our implementation\footnote{Our implementation is publicly available at \url{https://github.com/TU-Berlin-SNET/DIMS-openid-ssi_login}}, as with Sovrin, we do not have to rely on a difficult to verify central trust engine for evaluating the creditworthiness of a claim, as the Sovrin foundation takes the necessary steps to ensure that all Sovrin issuers are legally entitled to issue the credentials they are offering. However, Gr\"{u}ner et al.\ address interoperability by introducing an SSI broker component, which handles the complexity of interacting with various platforms, like Jolocom and uPort.

\subsection{Blockchain-secured Public Key Infrastructure}

Publications, patents, and whitepapers also exist that describe blockchain and distributed ledger applications for managing X.509 certificates.

Won \cite{won2018decentralized} imagines the implementation of a blockchain-based PKI, which enables faster certificate verification compared to the current state-of-the-art using X.509 certificates issued by certificate authorities for managing public keys for IoT devices. Won argues that the current PKI infrastructure is not suitable for IoT devices because (1) the consequences of failures in terms of security and (2) the difficulties with governing certificates due to the lack of standard protocols for managing the life cycle of public keys. Won argues that if the secret keys of a root CA are compromised by an attacker, then all entities must update their trusted root certificate lists accordingly. This was the case, for example, with the DigiNotar hack in 2011 \cite{digiNotar}.

Certificate Transparency\footnote{\url{https://www.certificate-transparency.org/}} (CT) provides a mechanism for reducing these risks by extending the current CA and PKI architecture. CT proposes a concept that combines three public services, consisting of a certificate log, log monitoring, and certificate auditing. These public services can run in a decentralized fashion by multiple providers, e.g., CDNs, ISPs, browser vendors, or DNS providers. Since 2018, browsers and search engines started to include additional certificate verification mechanisms, such as CT. Stark et al.\ show in \cite{Stark2019} that CT has been significantly adapted, but they also point out the danger that end users react incorrectly to, e.g., system warnings and thus the added value of CT is at risk. Another mechanism to strengthen the current CA infrastructure is the introduction of Extended Validation\footnote{\url{https://cabforum.org/extended-validation-2/}} (EV) certificates which demands the verification the entity's legal identity prior to granting a certificate. Only a subset of the current CAs are allowed to implement this approach.

Won \cite{won2018decentralized} further explains that Online Certificate Status Protocol\footnote{\url{https://tools.ietf.org/html/rfc2560}} responders can easily be overloaded due to the computationally intensive nature of generating responses, thus it is relatively easy to carry out successful distributed denial-of-service attacks on them. Won aims to replace CAs with a blockchain network based on Emercoin\footnote{\url{https://emercoin.com/enm}} utilizing a Proof-of-Stake consensus algorithm.

Originally, Blockstack \cite{AliBlockstackGlobalnaming2016} envisioned a global naming and storage system using the Bitcoin blockchain. In 2016, Blockstack served as a PKI system using Namecoin for its 40,000 users with over 33,000 entries and 200,000 transactions.

Blockstack PBC has then filed a patent application \cite{AliDecentralizedprocessingglobal2017} for a decentralized global naming system using a distributed hash table for realizing a virtual blockchain serving as a tamperproof registry for a global naming system.

DIDs can also be married with a DNS like naming system. The draft version of a DID method specification for Web\footnote{\url{https://w3c-ccg.github.io/did-method-web/}} is already available, which empowers DIDs by linking them to meaningful information behind already existing domains, thus helping mass adoption of the technology. An example is:

\begin{center}
\texttt{did:web:w3c-ccg.github.io}.    
\end{center}

\section{Concept and Design}
\label{sec:conceptdesign}

\subsection{Integrating the OpenID Connect standard claims with self-sovereign identity}

As there is a predefined set of claims\footnote{\url{https://openid.net/specs/}} for OIDC (see Table \ref{tab:oidc_claims}, all of them are strings, except for \emph{*\_verified} which are booleans, the \emph{address} which is a JSON object, and the \emph{updated\_at} which is a number), we should include these attributes into the OIDC schema and credential definitions. The claims include an identifier named \texttt{sub}, and several personal claims such as \texttt{name}, \texttt{gender}, \texttt{phone\_number}, \texttt{email}, \texttt{picture}, \texttt{website}, and \texttt{address}, \texttt{birthdate}. 

\begin{table}[b]
	\centering
	\caption{List of OIDC standard claims.}
	\label{tab:oidc_claims}
	\begin{tabular}{@{}lll@{}}
		\toprule
		sub & name & given\_name \\ \hdashline[0.5pt/2.5pt]
        family\_name & middle\_name & nickname \\ \hdashline[0.5pt/2.5pt]
        profile & picture & website \\ \hdashline[0.5pt/2.5pt]
        email & email\_verified & gender \\ \hdashline[0.5pt/2.5pt]
        birthdate & zoneinfo & locale \\ \hdashline[0.5pt/2.5pt]
        phone\_number & phone\_number\_verified & address \\ \hdashline[0.5pt/2.5pt]
        updated\_at \\ \bottomrule
	\end{tabular}
\end{table}

Phone number and email can be verified. As phone numbers are in general linked to physical identities, we could use them for trusted identity verification.

For supporting the OIDC standard claim set in a trusted manner, all we need with Hyperledger Indy is a \texttt{SCHEMA} and \texttt{CLAIM\_DEF} including all attributes. In this scenario, the end user does not even need an account at the SSI OIDC Provider, and would be able to log in with credentials from any issuer, e.g., via using the Sovrin network or others.

\subsection{DID Auth with OpenID Connect}

For authorization with our SSI-powered OIDC provider, we can rely on an \textit{authcrypted}, i.e., a public key authenticated encrypted and DID-signed message. Thus, we propose an additional VC-based authentication factor, where the requested VCs contain the necessary user information. There are three main benefits for the aforementioned process. First, the user data sent to the relying party is retrieved directly from the user. Second, we have the opportunity to gather additional information about our user. Third, VCs can establish more trust in user account information.

Receiving and verifying data from the holder in the form of a proof request would be a benefit from a trust, privacy, and flexibility perspective. If the verifier receives a proof, where the attributes are signed by a specific government, she could be virtually sure that the holder's real identity is revealed. Yet, the holder would also have the chance to send self-signed credentials for privacy reasons. Furthermore, this would also enable any Sovrin issuer to be part of the SSI OIDC Provider, thus help us achieving the goal of decentralizing identity, and giving end users a real opportunity to choose their preferred carrier from a much larger number of candidates than they have today. Still, application developers would only need to integrate the SSI OIDC provider with their web services, and yet they would be able to support login with accounts from more than one IdP.

A design decision about DID authentication for OIDC is if we conduct the VC proof and DID exchanges in one or two steps. Note, that a single step VC exchange can only be done once the SSI OIDC Provider has already established a pairwise DID connection, i.e., unique DIDs and related keys are already exchanged, with the end user, or if both parties have a public DID written on the distributed ledger. Furthermore, the SSI OIDC Provider also needs to know the DID of the user beforehand, so the proof request can be sent to the holder. For the one-step proof exchange communication between the holder and verifier can be encrypted and decrypted with the public and private keys of their DIDs.

\subsubsection{Authentication via an SSI mobile wallet app}

We need three things when using Hyperledger Indy: (1) an established pairwise connection between the IdP (issuer) and the user (holder), (2) knowledge about matching VC types, and (3) knowing which user we are authenticating. A pairwise connection between the SSI IdP and the holder can be established via scanning a QR code displayed on the SSI IdP's website with an SSI smartphone wallet app. For starting the authentication, we could use HTTP cookies or a challenge response based authentication scheme via scanning a QR, provided that no previous session information about the DID of the user exists. Then the identity holder receives a proof request and a push notification and sends the proof to the SSI OIDC provider upon consent by using biometric identification functions of the smartphone. Then the IdP can forward the received attributes in the form of a JSON Web Token to the relying party.

\begin{figure*}[t!]
	\centering
	\includegraphics[width=0.8\textwidth]{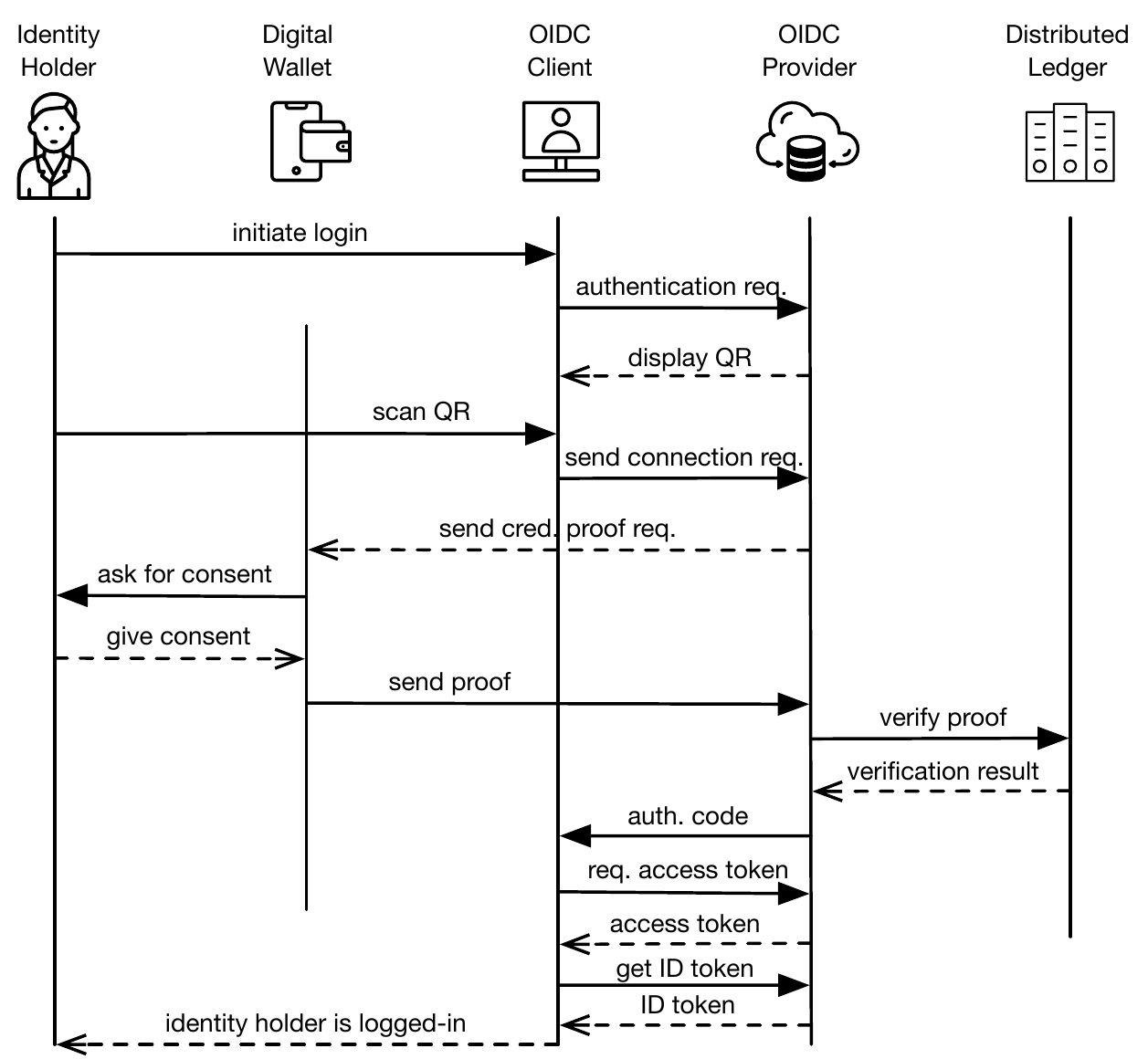}
	\caption{Sequence diagram of integrating OIDC-Authorization-Code-Flow-based SSO with SSI. In this variant, the DID authentication is performed in a single step with the proof exchange, where the sent proof contains the attributes for the id token in the form of a VC. First, the user initiates the login on a website. Then, the website (OIDC Client) redirects him/her to the SSI OIDC Provider. If the user does not yet have a connection, then this can be established by scanning a QR code with the identity wallet mobile app. The OIDC Provider then sends a proof request to the end user (identity holder). Next, the identity holder can decide whether to respond to the proof and thereby perform the login.}
	\label{fig:ssi_search}
\end{figure*}

\subsection{A distributed ledger-based Public Key Infrastructure and Domain Name System}

A public permissioned ledger like Sovrin is a promising candidate for a PKI, as the number of entities who need write access is orders of magnitude smaller than those with read access. The nodes running the consensus protocol need to be widely trusted organizations, for example the most trusted SSL certificate authorities (CA). Ordinary CAs can also have write access to the ledger. Mir-BFT \cite{stathakopoulou2019mir} is capable of ordering \textit{more than 60000 signed Bitcoin-sized transactions per second on a widely distributed 100 nodes}. Nowadays, web browsers rely on a predefined set of trusted root certificates. With a decentralized PKI, we could use the genesis transaction of a public permissioned ledger for initial verification of the authenticity of the retrieved certificates.

\subsubsection{NYM transaction-based PKI}
The creation of a DID known to the Ledger is an identity record itself, also called a NYM transaction. A NYM transaction can be used for creation of new DIDs that is known to a ledger. As NYM ledger transactions of Hyperledger Indy include both a \texttt{verkey} and an optional \texttt{alias} field, we can use them for binding a human readable name and a public key together for a specific DID. A \texttt{DID verkey} is a public key for an Edwards-curve Digital Signature Algorithm (EdDSA) signature scheme using SHA-512 hashing with the Ed25119 curve, i.e., an elliptic curve offering 128 bits of security, also suitable for generating X.509 certificates. Revocation of X.509 certificates can be supported by updating NYM transactions and the respective DIDs.

\subsubsection{Verifiable-credential-inspired PKI}
VCs are not stored on the public ledger but are held privately in the wallet of the identity holder with Hyperledger Indy. If websites would hold X.509-structured VCs with revocation support privately, we would not necessarily improve the transparency of the current PKI system, but we would still have a standardized way for obtaining the public keys of certificate authorities. Nevertheless, public access to proofs of VCs containing the same information as X.509 certificates is not an absolute must. It would be sufficient to have the public keys of certificate authorities on the Indy Ledger in NYM transactions, creating credential definitions recorded in \texttt{CLAIM\_DEF} transactions based on a X.509 certificate \texttt{SCHEMA} without making proofs publicly available.

\section{Implementation}
\label{sec:implementation}

We have a proof-of-concept implementation of the SSI OIDC Provider, and we have identified essential libraries for realizing the VC-based PKI.

\subsection{OpenID Connect provider with verifiable credential-based DID Auth}
For implementing the preferred SSI and OIDC integration we use the Indy Edge Agent API\footnote{\url{https://github.com/ID-Chain/IEA-API}}, an extended Node.js OIDC Provider, a test OIDC Client mocking the functionality of an online shop. Furthermore, a mobile SSI wallet application is also an important part of the workflow.
Figure \ref{fig:ssi_search} shows the implementation and the login process.
In the following, we describe each component.

\subsubsection{Indy edge agent API}

We have an own Indy edge agent REST API, that wraps the functionality of the Indy SDK, and is deployable on a server, where clients can use the functionality of the Indy SDK by using simple REST calls. Our Indy API can connect to any Indy ledger including the Sovrin MainNet. The API is written in Node.JS, and it also needs a MongoDB instance to serve as a persistent storage for identity wallets.

\subsubsection{SSI OpenID Connect provider}

For our proof of concept implementations of the SSI OIDC Provider we can base our solution on already existing, certified software. When working with OIDC we have several open source implementations to choose from.
We have selected a certified Node.js OIDC Provider\footnote{\url{https://github.com/panva/node-oidc-provider}} with a 100\% successful test matrix for token signing and verification on jwt.io, which is provided by the underlying JOSE library.

Our current implementation is using the implicit flow, but this can be changed relatively easily by configuring our selected OIDC Provider for the authorization code flow.

\subsubsection{SSI mobile wallet app}

We have developed a React Native Android app that serves as an SSI identity wallet for the identity holder using Hyperledger Indy. The React Native app relies on the Java wrapper for the Indy SDK. The app can establish pairwise connections with DIDs by scanning a connection offer QR code, and can acquire and store VCs. Furthermore, the app can also send proofs to verifiers upon receiving a proof request and giving consent from the identity holder.

\subsubsection{OpenID connect client}

In order to test our prototype, we have created a website, where the user logs in with his/her SSI credentials. The website then redirects the user to the SSI OIDC Provider, where the authentication is performed. If the process succeeds, then the client website receives the ID token. Currently, the implicit flow is used, where the client application receives the ID token as part of the redirected URL.
Currently, we use the Node.js OIDC Client with the implicit flow, which enables easy integration with OIDC Providers.

\subsection{Towards a distributed-ledger-powered PKI}

Attributes of an X.509 certificate can easily be incorporated in a VC as well instead of the Abstract Syntax Notation One (ASN.1) structure. In case of Hyperledger Indy, the \texttt{CLAIM\_DEF} would need to include most attributes in the certificate explicitly. The information found in the \texttt{Certificate\_Signature\_Algorithm} and \texttt{Certificate\_Signature} attributes would here be provided by the values found under the \texttt{reqSignature} field. In order to enable widespread usage of VC-based certificates we would have to integrate them with TLS and SSL libraries such as OpenSSL\footnote{\url{https://www.openssl.org/}} and Rustls\footnote{\url{https://docs.rs/rustls/0.17.0/rustls/}}.

\section{Evaluation}
\label{sec:evaluation}

There are various aspects for comparing blockchain and distributed-ledger-based authentication methods and centralized approaches for both the PKI and end user authentication scenarios.
In the following, we perform a qualitative evaluation.
Future work is to perform simulations and experiments regarding the real-world performance
of our methods.

\subsection{Distributed-ledger-based PKI}

As Won \cite{won2018decentralized} wrote, there is no standard protocol for managing the life cycle of X.509 certificates. A major advantage of a distributed-ledger-technologies-powered PKI system is more transparency and interoperability, where we could easily find all existing X.509 certificates and domain names on a public, permissioned, and tamper-proof registry.

A question to investigate is, if we are able to support the storage of all X.509 certificates on the ledger, or just the certificates of CAs. BuiltWith detected more than 145 million SSL certificates\footnote{\url{https://trends.builtwith.com/ssl/traffic/Entire-Internet}} as of March 2020. If we assume that all certificates will get renewed every 3 months, this would still result in an average transaction throughput of less then 17 transactions per second. If our distributed ledger supports 10,000 transactions per second, than we can expect our system to be able to record more than 124.5 million SSL certificates in less than 3.5 hours.

However, we still need more proof, research, and innovation to see if a distributes-ledger-based PKI can be a better alternative compared to the state-of-the-art. Nonetheless, the Sidetree protocol, platforms like Hyperledger Fabric, high throughput consensus algorithms like Mir-BFT \cite{stathakopoulou2019mir}, and the Sovrin architecture capable of serving a large volume of read requests by introducing observer nodes \cite{reed2016technical} all point in the direction that a distributed-ledger-based PKI system will soon become a viable solution at global scale.

\subsection{Self-sovereign OpenID Connect Provider}
We identify four major advantages of the SSI OIDC Provider, namely (1) interoperability, (2) privacy, (3) trust, and (4) richer client data.

First, it enables SSO with any Sovrin VC following the standard OIDC claims schema, yet the OIDC client only has to integrate with one OIDC Provider. This is a benefit for OIDC clients, as it is impractical for the user to manually select from a possibly high number of IdP, and it also reduces development efforts. For example, the client has to integrate with Google, GitHub, Microsoft, Yahoo, and Amazon logins individually, and then the user also has to pick his/her preferred choice. Integration overhead and the list of different login options to display for the user scale linearly in the number of OIDC Providers on the OIDC Client side. But with our SSI OIDC Provider, it would always just be one, yet the number of different accounts supported could be arbitrarily high.

Second, as end users get a proof request about the data they disclose, they have more control over what personal information they share, and the whole process becomes more transparent. Furthermore, the identity wallets of end users are stored by themselves on their preferred devices, and not by third parties. The SSI and OIDC integration protocol from DIF leveraging self-issued ID tokens offers the most possible privacy, however, it requires OIDC clients to support a special feature. Our proposed SSI OIDC provider does not require any non-default feature from OIDC clients.

Third, VCs with Sovrin come from trusted issuers, which are onboarded by the Sovrin foundation. As the Sovrin ledger is permissioned, schemas and credential definitions can be written on the Sovrin MainNet only by trustworthy entities. This guarantees that verifiers can trust Sovrin credentials.

Finally, regular OIDC Providers might not support all attributes of the standard claim set according to the OIDC specification. However, if such attributes would be backed by VCs, then an SSI OIDC provider could serve its clients with a rich claim set if the user owns matching credentials.

\section{Conclusion}
\label{sec:conclusion}

Decentralizing authentication with SSI benefits the end user in IoT, PKI, and OIDC scenarios. 

Marrying OIDC with SSI and retrieving the account claims as VCs not only gives more freedom to end users, but also enables login with a single SSI OIDC provider without an account by presenting any matching VC. Possible next steps for our SSI OIDC provider prototype would be providing support for login using VCs and DIDs from more SSI platforms like Jolocom and uPort, and not only from Sovrin and Hyperledger Indy.

The result of our analysis is, that in the IoT PKI scenario, the deployment of a permissioned ledger could serve as a publicly verifiable registry for the necessary cryptographic public data. Such data include X.509 certificate metadata needed for verifying verifiable credentials.

In conclusion, the SSI movement has already contributed several feasible variants for authentication with OIDC, including our proof-of-concept SSI OIDC provider. Furthermore, a distributed ledger-based PKI for IoT devices could technically become viable at scale.

\section*{Acknowledgment}
This activity has received funding from the European Institute of Innovation and Technology (EIT) under the project Decentralized Identity Management System (DIMS). This European body receives support from the Horizon 2020 research and innovation program. We would also like to thank {\"O}mer Ilhan, Santhos Baala Ramalingam Santhanakrishnan, and Marcel Ebermann for their support in the implementation of the prototypical components.

\bibliographystyle{IEEEtran}

\balance
\cleardoublepage
\includepdf[pages=-]{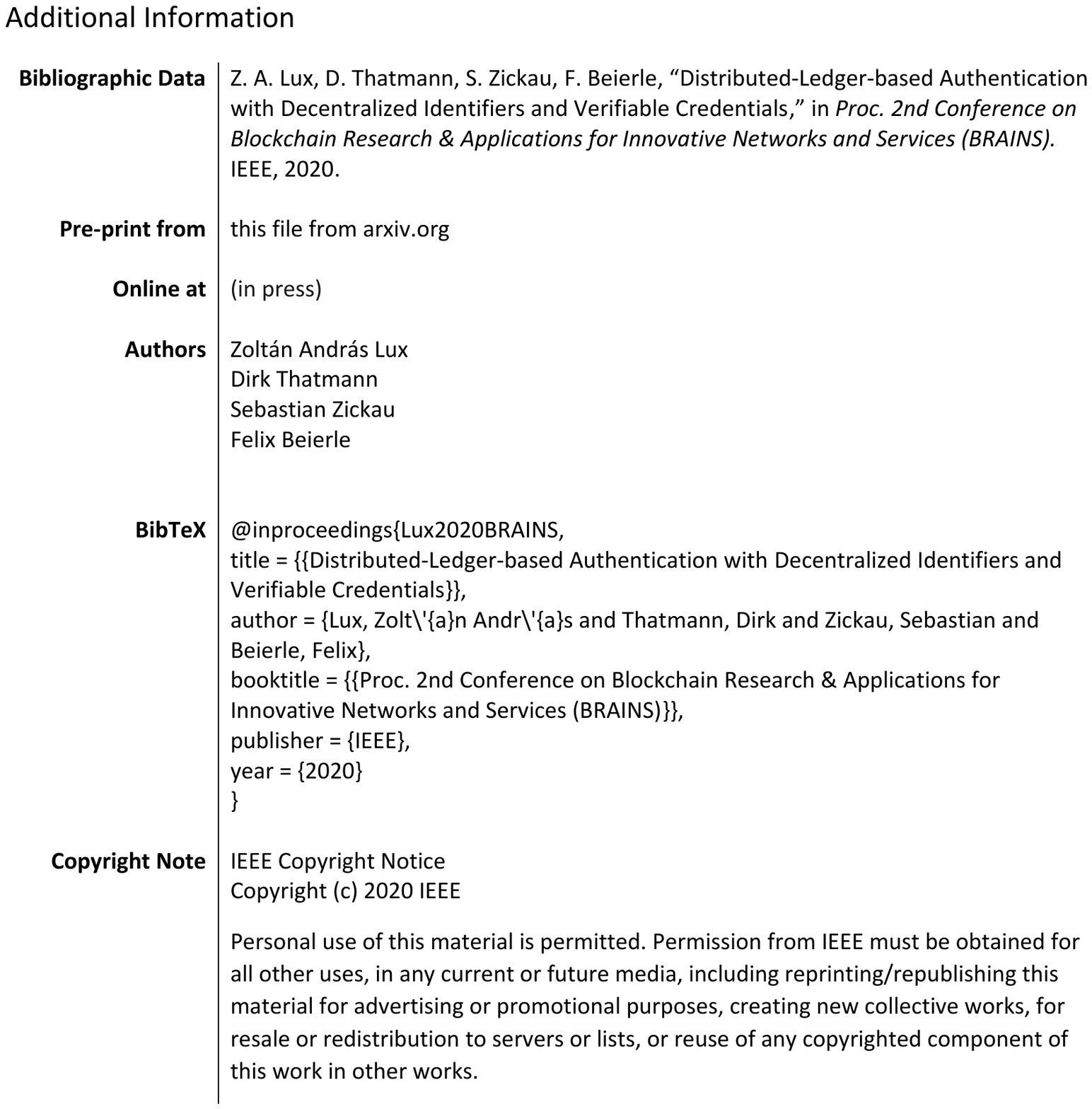}

\end{document}